\input{aipcheck}

\documentclass[final]{aipproc}
\layoutstyle{6x9}

\begin{document}

\title{Singularities around $w=-1$}


\author{Leonardo Fern\'andez-Jambrina}{
  address={ETSI Navales, Arco de la Victoria s/n, E-28040-Madrid, 
  Spain}
}

\begin{abstract}
In this talk we would like to analyse the appearance of singularities
in FLRW cosmological models which evolve close to $w=-1$, where $w$ is the
barotropic index of the universe.  We relate small terms in
cosmological time around $w=-1$ with the correspondent scale factor of
the universe and check for the formation of singularities.
\end{abstract}

\maketitle

%%%%%%%%%%%%%%%%%%%%%%%%%%%%%%%%%%%%%%%%%%%%
%% MAINMATTER
%%%%%%%%%%%%%%%%%%%%%%%%%%%%%%%%%%%%%%%%%%%%

\section{Introduction}

Accelerated expansion of our universe has been established
observationally in the last decade from several sources of
information.  Even more, the barotropic index $w$ of the equation of
state of the content of the universe, that is the ratio between
pressure and energy density, has been checked to be close to $-1$,
value which has been dubbed as the phantom divide.  This value
corresponds to a cosmological constant as energy content of the
universe.

The fact of the accelerated expansion of the universe has lead to 
several scenarios for a singular fate of the universe, different from 
eternal expansion or Big Crunch. A classification of such 
singularities is provided in \cite{Nojiri}. Among these new 
scenarios one can find Big Rip singularities \cite{Caldwell}, 
sudden singularities \cite{sudden}, Big Freeze singularities 
\cite{freeze}, $w$-singularities \cite{wsing} or directional 
singularities \cite{hidden}.

Some of then however have been shown to be not strong enough
\cite{puiseux} to mean the actual end of the universe. For instance, 
sudden singularities \cite{suddenferlaz} and $w$-singularities 
\cite{wchar} are weak singularities. They have also been studied 
within the framework of modified theories of gravitation 
\cite{modigravi}.

The main idea behind the previous analysis is to assume that the 
scale factor of the universe may be written as generalised power 
expansion of the time coordinate \cite{visser} and then analyse the 
geodesic completeness \cite{HE} of the FLRW spacetime endowed with such 
scale factor.

With these ideas in mind, we would like to analyse the behaviour of 
cosmological models in the vicinity of $w=-1$. We shall show that 
there are scenarios which fall out of the previous classification and 
are to be regarded separately.
\section{Perturbed barotropic index}

Let us assume that the barotropic index behaves approximately as a 
cosmological constant at present time, that is, 
\begin{equation}w=-1+h(t),\qquad |h(t)|\ll 1.\end{equation}

It is formally possible to get an explicit expression for the scale 
factor $a(t)=e^{f(t)}$ for such barotropic index,
\[
w=\frac{p}{\rho}=-\frac{1}{3}-\frac{2}{3}\frac{a\ddot a}{\dot a^2}
\Rightarrow f(t)=\frac{2}{3}\int 
\frac{dt}{\int h(t)\,dt+k}+C.\]

The constant $C$ is irrelevant since it amounts to a change of scale 
or time origin, $a(t)\rightarrow e^Ca(t)$. Since the density of the energy content is
\[\rho=\left(\frac{\dot a}{a}\right)^2=\dot f^2 \Rightarrow 
\sqrt{\rho}=\frac{2}{3\int h+k},\]
we may use it to fix the constant $k$.

For instance, taking the origin in the near future and assuming a 
power-law deviation from unity, $h(t)=\alpha (-t)^p$,  $p>0$ so that $w(0)=-1$,
\[\sqrt{\rho(t)}=-\frac{2(p+1)}{3\alpha}\frac{1}{k+(-t)^{p+1}}.\]

Two different cases arise in these models:

\begin{itemize}
    \item $k\neq0$: $\sqrt{\rho}$ has simple poles out of $t=0$ and 
    $\rho\sim (t-t_{0})^{-2}$, but $w=-1$ is regular.

    \item $k=0$: Singular density and scale factor at $w=-1$. Naming 
    $\beta=2(1+p)/3\alpha p$,
\[\rho(t)=\frac{4}{9}\left(\frac{p+1}{\alpha}\right)^2\frac{1}{t^{2p+2}},\quad f(t)=-\frac{2}{3}\frac{1+p}{\alpha p}\,(-t)^{-p}, \qquad
a(t)=e^{-\beta/(-t)^p}.\]
\end{itemize}

We are interested in the latter models, which have a singular density 
at the time when $w=-1$ is reached. These models have a 
non-analytical scale factor with an essential singularity at $t=0$ 
and the energy density blow up as $1/t^{2p+2}$ instead of 
$t^{-2}$, which is the usual power for divergencies with analytical 
scale factors. Two possibilities arise depending on the sign of the 
constant $\alpha$:

\begin{itemize}
   \item For $\alpha>0$, we have a \emph{Great Crunch}: 
   $a(t)\rightarrow 0$.
   
   \item For $\alpha<0$, we have a \emph{Great Rip}: 
    $a(t)\rightarrow \infty$.
\end{itemize}
\section{Geodesics near $w=-1$}
Equations for causal geodesics parametrised by proper time $\tau$,
\[d\tau=\sqrt{-g_{ij}dx^idx^j},\] in a flat FLRW model reduce to
\[\frac{dt}{d\tau}=\sqrt{\delta +\frac{P^2}{a^2(t)}},\qquad 
\frac{dr}{d\tau}=\pm\frac {P}{a^2(t)},\]where $P$ is a constant of 
motion and $\delta=0$ for null and 
   $\delta=1$ for timelike geodesics.

The case of null geodesics is simpler since the equations may be 
integrated, \[\tau=P^{-1}\displaystyle\int_{t_{0}}^{0}a(t)\,dt.\]

This means it takes an infinite proper time to Great Rip 
($a(0)=\infty$), so this singularity is not accesible along lightlike 
geodesics.

On the contrary, for timelike geodesics,
\[\tau=\displaystyle\int_{t_{0}}^{0}\frac{dt}{\sqrt{1 
+P^2a^{-2}(t)}},\]
all geodesics reach $w=-1$ in finite proper time.

Hence $w=-1$ becomes singular but for null geodesics. This sort of 
behaviour is similar to the one in the vicinity of a Big Rip 
singularity.

\section{Strong singularities}
In spite of the singular character of $w=-1$, it could happen that 
the singularity could be not strong enough to be the end of the 
universe, since  extended objects might avoid 
being disrupted by tidal forces on crossing the singularity.  

This concept of strength of singularities was first coined by Ellis
and Schmidt \cite{HE} and it was related to the cosmic censorship
conjecture later on.  It was further developed by other authors.  For
instance, for Tipler \cite{tipler} a singularity is strong if the
volume spanned by three Jacobi fields orthogonal to the velocity of
the geodesic tends to zero as it approaches its end.  Kr\'olak
\cite{krolak} suggests a less restrictive criterion, requiring just the derivative of
the volume with respect to proper time to be negative.

Since these definitions involve calculations with Jacobi fields, 
checking the strength of singularities may be cumbersome. 
Fortunately, necessary and sufficient
conditions \cite{clarke} have been derived involving just integrals of some 
components of the curvature tensor along incomplete geodesics.

For instance, following Tipler's definition, a singularity is strong
along a null geodesic of velocity $u$ if and only if the
integral
\[\displaystyle
\int_{0}^{\tau}d\tau'\int_{0}^{\tau'}d\tau''R_{ij}u^{i}u^j\] diverges
as $\tau$ tends to $\tau_{0}$, where $\tau_{0}$ is the proper time 
assigned to the singularity.

For Kr\'olak's criterion the necessary and sufficient condition is 
the divergence of the integral
\[\displaystyle \int_{0}^{\tau}d\tau'R_{ij}u^{i}u^j.\]
as $\tau$ tends to $\tau_{0}$.

Such calculations are simple in our case and allow us to conclude 
that singularities at $w=-1$ are strong according to both criteria. 
The same result is obtained for timelike geodesics.
 \section{Conclusions}
We have shown that there are two possible behaviours for models with 
a barotropic index close to $w=-1$:
\begin{itemize}
    \item  Regular crossing of $w=-1$.

    \item  Essential singularity at $w=-1$: Great Crunch / Rip.
\end{itemize}

For the latter models the energy density blows up at the singularity
and its divergence is worse than $t^{-2}$.  The essential
singularities are strong, though null geodesics never reach the Great 
Rip. More details and references may be found elsewhere \cite{paper}.
\section*{Acknowledgments}The author wishes to thank the University of
the Basque Country for their hospitality and facilities to carry out
this work.

\end{document}